%% file: zurek-spotlight.tex
\documentclass[aip,jcp,twocolumn,10pt]{revtex4-1}
\usepackage{times}
\usepackage{amsmath}
\usepackage{amsfonts}
\usepackage{graphics,graphicx}

\usepackage{color}
\usepackage{tabularx}
\usepackage[T1]{fontenc}
\begin{document}  %% Titlepage
%%%%%%%%%%%%%%%%%%%%%%%%%%%%%%%%%%%%%%%%%%%%%%%%%%%%%%%%%%%%%%%%%%%%%%%%%%%%%%%

\title{High-Temperature Superconductivity in Alkaline and Rare Earth Polyhydrides at High Pressure: A Theoretical Perspective}

\author{Eva Zurek}\email{ezurek@buffalo.edu}
\affiliation{Department of Chemistry, State University of New York at Buffalo, Buffalo, NY 14260-3000, USA}
\author{Tiange Bi}
\affiliation{Department of Chemistry, State University of New York at Buffalo, Buffalo, NY 14260-3000, USA}
\begin{abstract}
The theoretical exploration of the phase diagrams of binary hydrides under pressure using \emph{ab initio} crystal structure prediction techniques coupled with first-principles calculations has led to the \emph{in silico} discovery of numerous novel superconducting materials. This Perspective article focuses on the alkaline earth and rare earth polyhydrides whose superconducting critical temperature, $T_c$, was predicted to be above the boiling point of liquid nitrogen. After providing a brief overview of the computational protocol used to predict the structures of stable and metastable hydrides under pressure, we outline the equations that can be employed to estimate $T_c$. The systems with a high $T_c$ can be classified according to the motifs found in their hydrogenic lattices. The highest $T_c$s are found for cages that are reminiscent of clathrates, and the lowest for systems that contain atomic and molecular hydrogen. A wide variety of hydrogenic motifs including 1- and 2-dimensional lattices, as well as H$_{10}^{\delta-}$ molecular units comprised of fused H$_5^{\delta-}$ pentagons are present in phases with intermediate $T_c$s. Some of these phases are predicted to be superconducting at room temperature. Some may have recently been synthesized in diamond anvil cells.

\end{abstract}

\maketitle

\section{Introduction}
The pressure induced metallization of hydrogen was first proposed by J.\ D.\ Bernal, but only later reported in the literature by Wigner and Huntington, who  predicted that at $P>$~25~GPa hydrogen would become an alkali metal-like monoatomic solid \cite{Wigner:1935}. It turned out that this was a gross underestimate: the lightest element has stubbornly resisted metallization up to center of the Earth pressures in a diamond anvil cell (DAC) \cite{Narayana:1998a,Loubeyre:2002,Zha:2013a,Eremets:2011a,Howie:2012a,Simpson:2016a,Zha:2012a,Howie:2012c,Goncharov:2013a}. A few recent studies have presented evidence indicative of metallization \cite{Eremets:arxiv-2016, Dias:2017a}, but the interpretation of the experimental results has been questioned \cite{Loubeyre:arxiv-2017,Eremets:arxiv-2017,Liu:2017a}. 

Why has blood, sweat and tears gone into attempts to metallize hydrogen? And what makes metallic hydrogen so special that it has been dubbed \emph{the holy grail of high pressure research}? The reason can be traced back to Ashcroft, who predicted that the large phonon frequencies  (a result of the small mass of H), large electron phonon coupling (arising from the strong covalent bonds between the H atoms, and lack of core electrons), wide bands, and substantial density of states (DOS) at the Fermi level ($E_F$) would render this elusive substance superconducting at high temperatures \cite{Ashcroft:1968a}. 
 Indeed, first principles calculations have estimated high values for the superconducting critical temperature ($T_c$): 242~K at 450~GPa in the molecular phase \cite{Cudazzo:2008a}, and as high as 764~K for monoatomic hydrogen near 2~TPa \cite{Mcmahon:2011b}. Thus, one way to make superconducting metallic hydrogen is by achieving the appropriate $T$/$P$ conditions.

\begin{figure}
\begin{center}
\includegraphics[width=1\columnwidth]{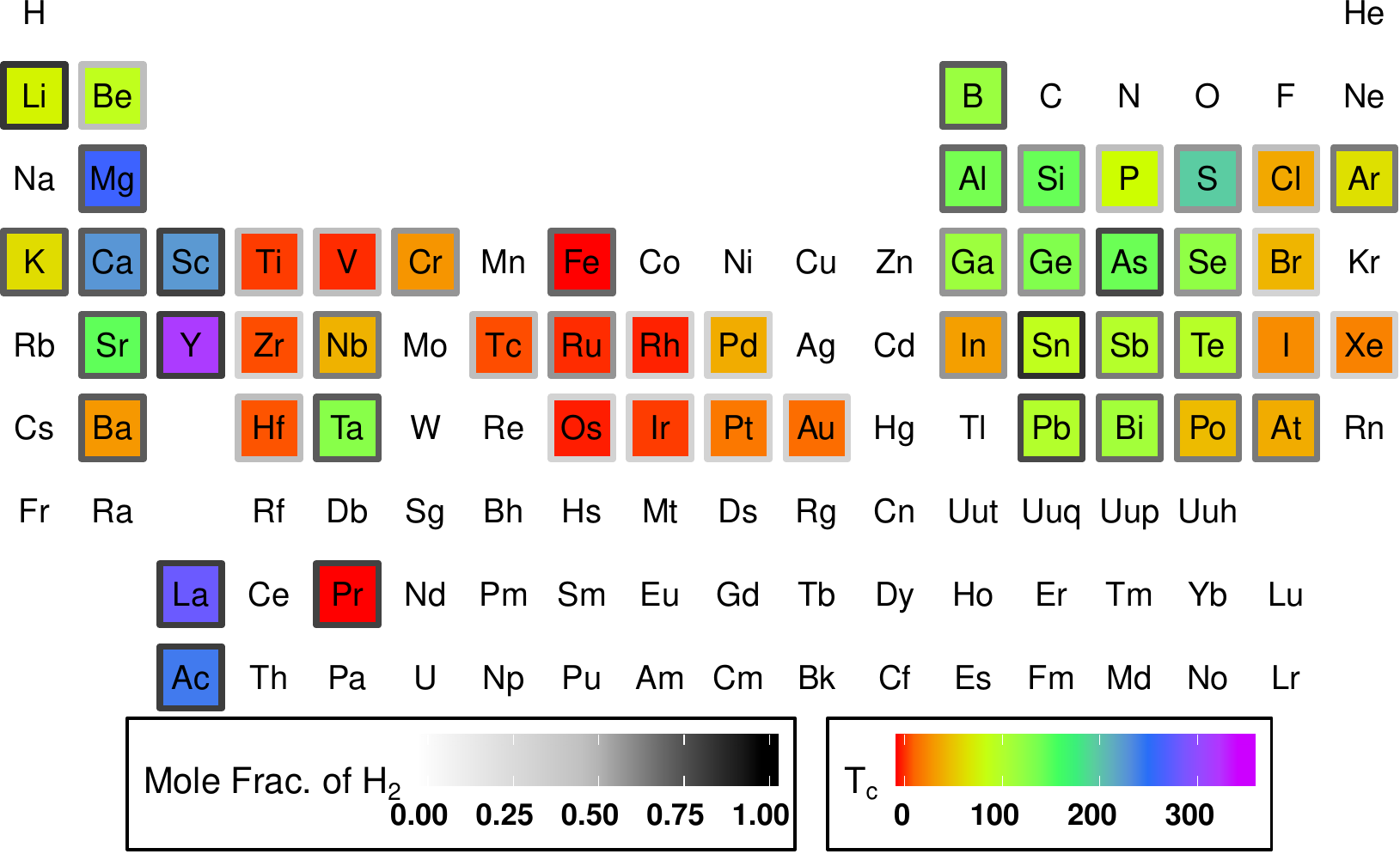}
\end{center}
\caption{The periodic table colored according to the highest $T_c$ value theoretically predicted for a binary hydride containing the given element \cite{note3}. The coloring scheme, in Kelvin, is given below. The H$_2$ mole fraction in the hydride with the highest $T_c$ is represented by the color outlining the element. It ranges from white (low) to black (high). If the element is not enclosed in a colored box, then theoretical predictions of the $T_c$s of its hydrides are not available. Typically this is because the phases predicted to be stable are not metallic or are poor metals, and they are therefore unlikely to have a high $T_c$.}
\label{fig:periodic-table}
\end{figure}

However, \emph{chemistry} may be able to hasten the metallization of hydrogen. In a seminal manuscript Ashcroft pointed out that because the hydrogen in solids comprised of the group 14 hydrides is ``chemically precompressed'', they may become metallic at lower pressures than elemental hydrogen \cite{Ashcroft:2004a,Ashcroft:2004b}. In other words, hydrogen's interaction with the tetragen atoms would modify the H-H distances and densities in such a way that less external physical pressure would be required to achieve a metallic state. 
Moreover, Ashcroft hypothesized that these compressed hydrides would have the same properties conducive towards high temperature superconductivity as hydrogen itself, thereby suggesting a parallel route to metallizing hydrogen. These predictions led to a plethora of theoretical, and a few experimental studies that searched for superconductivity in hydrides known to exist at 1~atm, including SiH$_4$ \cite{Feng:2006a,Pickard:2006a,Yao:2007-Si,Degtyareva:2007-Si, Chen:2008-2-Si,Kim:2008-Si,Martinez-Canales:2009-Si,Yan:2010-Si,Zhang:2015-Si,Cui:2015-Si,Eremets:2008a}, GeH$_4$ \cite{Canales:2006a,Li:2007-Ge,Gao:2008a,Zhang:2010-Ge,Zhang:2015-Ge}, and AlH$_3$ \cite{Pickard:2007b,Goncharenko:2008-Al,Kim:2008-Al,Geshi:2013-Al}. But, because pressure is known to affect the chemical compositions that are stable \cite{Zurek:2014i,Zurek:2016b,Hermann-lip}, systems with stoichiometries that are not observed at 1~atm, e.g.\ SiH$_4$(H$_2$)$_n$, $n>0$, \cite{Strobel:2009a,Wang:2009-Si,Yao:2010-Si,Yim-2010-Si,Michel:2010-Si,Shanavas:2012-Si,Li:2010-Si}, and LiH$_n$, $n>1$, \cite{Zurek:2009c,Pepin:2015a} were also investigated. The possibility of accessing novel combinations and electronic structures under pressure dramatically widens the phase space wherein a high temperature superconductor can be found.

High pressure experiments can be very costly, and the results can be difficult to analyze. It is therefore no surprise that the dramatic improvement of computer hardware, coupled with the development of robust methods for \emph{a priori} crystal structure prediction (CSP) using Density Functional 
Theory (DFT) calculations has led to the rapid \emph{in silico} exploration of the phase diagrams of materials under pressure.
Many investigations have focused on the parallel route of metallizing hydrogen inspired by Ashcroft's predictions \cite{Ashcroft:2004a,Ashcroft:2004b}.
In just over a decade the phase diagrams of most binary hydrides under pressure have been calculated \cite{Zurek:2016d,Shamp:2016,Zhang:2017,Duan:2017a,Wang:2017a,Struzhkin:2015a,random3,Yao-S-review:2018,Zurek:2018d,Semenok}. $T_c$s that approach, and even surpass, room temperature have been calculated for a number of phases that are predicted to be stable at pressures attainable in DACs. 

The periodic table illustrated in Fig.\ \ref{fig:periodic-table} graphically shows the highest undisputed \cite{note2} $T_c$s predicted for each element's polyhydride \cite{note3}. Two regions contain hydrides that are the most promising for high temperature superconductivity: (i) many of the $p$-block elements in groups 13-16, and (ii) a subset of the alkaline earth and rare earth metals. High temperature superconductivity was found within the first region when a $T_c$ of 203~K near 150~GPa was measured in a sample of compressed hydrogen sulfide \cite{Drozdov:2015a}. Not only does this material possess the highest confirmed $T_c$ to date, remarkably it is also a Bardeen-Cooper-Schrieffer (BCS) type superconductor \cite{Mazin:2015a}. As described in more detail elsewhere \cite{Yao-S-review:2018}, this discovery was not serendipitous; theoretical investigations inspired \cite{Li:2014} experiment and helped to interpret the results \cite{Duan:2014}. Superconductivity in the sulphur/hydrogen system under pressure has been the topic of numerous studies \cite{Einaga:2016,Troyan:2016-S,Capitani:2017a,Goncharov:2017,Guigue:2017,Flores:2016a,Papaconstantopoulos:2015,Bernstein:2015,Duan:2015-S,Errea:2015a,Akashi:2015-S,Errea:2016,Quan:2016a,Sano:2016a,Ortenzi:2016a,Gorkov:2016a,Goncharov:2016a,Bussmann:2017-S,Durajski:2016-S-P,Jarlborg:2016a,Szczkesniak:2017-S,Azadi:2017,Durajski:2015-S,Arita:2017a,Gordon:2016,Majumdar:S-2017,Akashi:2016a,Li:2016-S,Ishikawa:2016,Heil:2015a,Ge:2016-S,Bianconi:2015a}, and it has been reviewed \cite{Yao-S-review:2018,Eremets:2016-review,Einaga:2017}.

This Perspective focuses on studies of hydrides in the second region, the alkaline and rare earths, whose predicted $T_c$s are at least as high as the boiling point of liquid nitrogen. Recent experiments tantalize with the promise of room temperature superconductivity in these systems, with measured $T_c$s as high as 215~K \cite{Drozdov:2018-La} or 260-280~K \cite{Zulu:2018-La} in the lanthanum/hydrogen system under pressure. Sec.\ \ref{sec:comp} provides a brief overview of the computational approaches used for the \emph{ab initio} prediction of the structures of novel pressure stabilized hydrides, and Sec.\ \ref{sec:tc} discusses the equations used to approximate $T_c$. The predicted structures and their superconducting properties are discussed in Sec.\ \ref{sec:motifs}, which is organized by the motifs found in these phases' hydrogenic lattices: clathrate-like arrangements, mixed atomic and molecular hydrogen, and other unique discrete or periodic hydrogenic motifs, such as H$_5^{\delta-}$ ``pentagons'' or 2-dimensional sheets. A brief outlook is provided in Sec.\ \ref{sec:outlook}.

\section{Computational and Theoretical Considerations} \label{sec:comp}

Pressure can have a profound effect on the compositions, structures, properties and stability of solid phases \cite{Hemley:2000,Song:2013a,Grochala:2007a,Goncharov:2013b,Bhardwaj:2012a,Dubrovinsky:2013a,McMillan:2013a,Klug:2011a,Naumov:2014a,Zurek:2014i,Zurek:2016b,Hermann-lip}. Who would have guessed that the ``simple metal'' sodium becomes an insulator by 200~GPa \cite{Ma:2009a}, the ``noble'' gas helium reacts with sodium to form a stable Na$_2$He \cite{Dong:2017a,Zurek:2017i} phase at 113~GPa, and hydrides with unusual stoichiometries such as NaH$_3$ \cite{Struzhkin:2016}, Xe(H$_2$)$_8$ \cite{Somayazulu:2010a} and FeH$_5$ \cite{Pepin:2017a} can be synthesized in a DAC? Such chemistry could not have been predicted using rules and bonding schemes that are based upon our experience at 1~atm. At the same time, experiments that approach the pressures in the Earth's core ($\sim$350~GPa) are challenging, and the results can be difficult to analyze. This has led to the development of a symbiotic relationship between theoreticians and experimentalists, where a feedback loop between experiment and theory has often led to important discoveries. 

Because of the difficulty inherent in performing high pressure syntheses and fully characterizing the phases made, many theoretical groups have employed \emph{a priori} methods for crystal structure prediction (CSP) to propose stable candidate structures. Structure prediction is a global optimization problem, where the goal is to determine the unit cell parameters, and atomic coordinates that correspond to the global minimum, as well as low lying local minima, in the potential energy surface (PES). However, as with all optimization problems, it is not possible to guarantee that the global minimum has been found within a CSP search. Thus, many of the algorithms that have been adapted towards CSP employ well known meta-heuristics that can find sufficiently good solutions for the minima. This includes: (quasi) random searches, techniques based on swarm behavior, evolutionary (genetic) algorithms, basin or minima hopping, as well as simulated annealing and metadynamics. A number of excellent reviews describing these methods \cite{woodley:2008a,Schon:2010a,Rossi:2009a,Wang:2014a,random1,revard,oganov:2010a,wales:2003a,Zurek:2014d,uspex5,Jansen:2015a}, and the synergy between CSP and experiment in high pressure research  \cite{Zurek:2014i,Zhang:2017,Wang:2014a,random1,random3} are available.

\begin{figure*}
\begin{center}
\includegraphics[width=2.0\columnwidth]{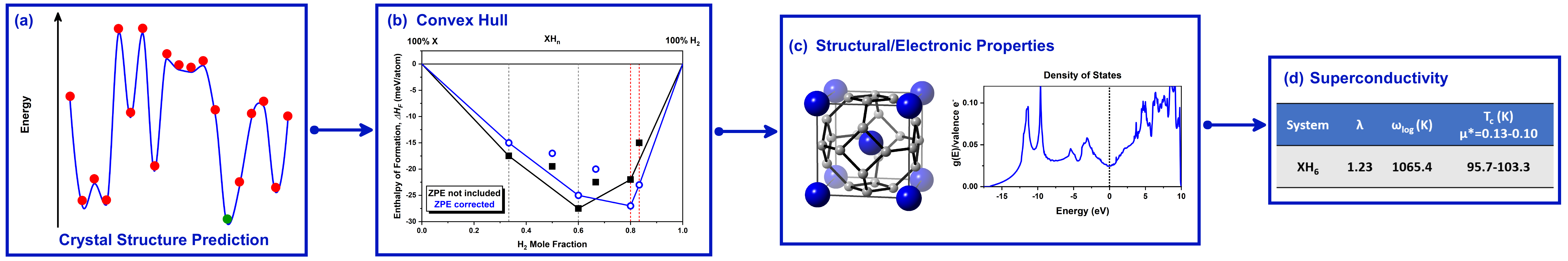}
\end{center}
\caption{Typical workflow used to predict pressure stabilized hydride based superconductors. (a) Candidate structures are found using \emph{ab initio} crystal structure prediction techniques. (b) The $\Delta H_F$ of the most promising phases are plotted vs.\ the mole fraction of H$_2$ they contain. The convex hulls without, and with ZPE contributions are determined. (c) The electronic structure of thermodynamically stable and select metastable species are computed to identify systems that have a high DOS at $E_F$. (d) Electron phonon coupling calculations are carried out, and the results are used to estimate $T_c$ via the equations given in Sec.\ \ref{sec:tc}.}
\label{fig:flowchart}
\end{figure*}
Typically hundreds, if not thousands, of structures are optimized to the nearest local minimum within a single CSP search. Moreover, often multiple searches need to be carried out for different stoichiometries and for a range of pressures. Because computations of the vibrational contributions to the free energy are significantly more time consuming than geometry optimizations, most CSP searches explore the 0~K PES. Reliable force fields are often not available for matter at extreme conditions, so the local optimizations are usually performed using DFT with a functional based on the generalized gradient approximation (GGA); often the Perdew, Burke and Enzerhof (PBE) \cite{Perdew:1996a} GGA is employed. Hybrid functionals such as HSE06 \cite{Krukau:2006a} or PBE0 \cite{Adamo:1999a} have so far not been used during CSP because of the immense computational expense involved. However, the inclusion of Hartree-Fock exchange can significantly affect the calculated transition pressures between different phases, \cite{Liu:2012aaa,Teweldeberhan:2012a,Calle:2015a}, stabilization pressures \cite{dominik:2016a,dominik:2016b,Hermann:2016-Wojciech}, computed $T_c$ values \cite{Komelj:2015a}, and it can lead to symmetry breaking structural (Peierls) distortions \cite{Boates:2011a}. Even though still rare, it is becoming more common to reoptimize the most stable structures found in a GGA-CSP search using hybrid functionals \cite{dominik:2016a,dominik:2016b,Hermann:2016-Wojciech}. In situations where metallization occurs because of pressure induced band broadening of the valence and conduction bands, and their eventual overlap, GGA functionals predict a too low metallization pressure. Hybrid functionals provide better estimates of the pressure at which band gap closure occurs. The inclusion of dispersion can also affect transition pressures for molecular solids \cite{Santra:2011a,Santra:2013b}, and DFT+U has been employed for strongly correlated systems \cite{Kurzydlowski:2018a}. An \emph{adaptive} genetic algorithm, which uses DFT calculations performed on a small set of candidate structures to parametrize, on-the-fly, classical potentials that are then employed for the local minimizations has successfully been applied to predict the structures of binary and ternary phases under pressure \cite{adaptive:2014a}.

For systems containing light elements the inclusion of the zero point energy (ZPE) can influence the relative enthalpies of different phases, as well as the identity of the thermodynamically stable species and their stability ranges. This is especially true if the hydrogenic sublattices of the structures whose enthalpies are compared have different motifs. Cold melting, which can occur if the difference in enthalpy between different phases is smaller than the ZPE, has been proposed for metal hydrides that contain a heavy element \cite{Hooper:2011a,Zaleski-Ejgierd:2011a}. It is currently standard practice to consider how the ZPE affects the stability of the phases found in a CSP search carried out at 0~K, but finite temperature contributions to the free energy are rarely considered, even though they may also be important (as it has been shown for elemental hydrogen \cite{Pickard:2012a}). 
Because the experimental conditions employed can affect which phases (i.e.\ stable or metastable) are made \cite{Zurek:2018c,Drozdov:2015-P}, it is important for theory to identify not only the global minimum, but also low lying local minima \cite{Zurek:2018h}.

Fig.\ \ref{fig:flowchart} illustrates the typical workflow employed by theoreticians searching for novel pressure-stabilized superconducting phases, including those discussed in this Perspective. First, CSP searches are carried out on the 0~K GGA PES as a function of composition and pressure. Promising candidate structures are optimized using more accurate settings, and their enthalpies of formation, $\Delta H_F$, are plotted vs.\ the mole fraction of H$_2$ they contain. The $\Delta H_F$ are employed to generate the convex hull, which is the set of line segments below which no other points lie. The phases whose $\Delta H_F$ comprise the hull are thermodynamically stable. Phonon calculations are carried out to confirm dynamic stability of phases that are on the hull or close to it, and the results are used to construct a new convex hull that includes the ZPE contributions to the enthalpy at 0~K. The structure, and electronic structure of select systems are computed and analyzed. In the following Section we described the methods employed to estimate the $T_c$ of the most promising structures. 

All of the phases discussed in the following sections were predicted using particle swarm optimization \cite{Wang:2010a,calypso}, evolutionary algorithms \cite{uspex1,Zurek:2011a} or \emph{ab initio} random structure searching \cite{random2}.
Because they did not become metallic because of pressure induced band broadening, it was not necessary to recompute their DOS' with a hybrid functional. Even though it is possible that geometry optimization with a hybrid functional would result in a Peierls distortion that opens up a band gap, this was not considered. Because of the high computational cost associated with performing electron phonon coupling calculations, $T_c$ was estimated using GGA calculations within the harmonic approximation.

\section{Estimating the Superconducting Critical Temperature} \label{sec:tc}
In this section the equations that are typically employed to estimate $T_c$ for compressed hydrides from first principles calculations are introduced. In the past these techniques have been successfully applied to a wide range of metallic systems \cite{bose2009electron}. If the superconducting properties of a material can be described using BCS theory, or its extensions, the material is referred to as a conventional superconductor. The electron phonon coupling mechanism that leads to superconductivity within BCS theory can be pictured as follows. When an electron passes through a crystalline lattice of positively charged ions, the Coulomb attraction between the electron and the ion causes the lattice to distort. The ionic displacement, in turn, attracts another electron with opposite spin and momentum. The two electrons become correlated and form a boson-like quasiparticle called a Cooper pair. All of the Cooper pairs in a superconductor break up at a temperature higher than $T_c$, thereby destroying the superconducting state.

Within BCS theory, $T_c$ can be estimated via
\begin{equation}
 T_c = 1.14 \langle{\omega}\rangle \exp\left[\frac{-1}{g(E_F)V}\right]
\end{equation}
where $\langle{\omega}\rangle$ is the average phonon energy, $g(E_F)$ is the single spin electronic DOS at $E_F$, and $V$ is the pairing potential between two electrons that occurs via the electron phonon interaction (which is assumed to be constant within  $2\hbar\omega_\text{cut}$ of the Fermi surface, and zero otherwise) \cite{bcs1,bcs2}. The cutoff frequency, $\omega_\text{cut}$, is often taken to be the Debye frequency or the average frequency, and it defines the size of the superconducting band gap.

Despite the tremendous successes of BCS theory in describing the properties of conventional superconductors, substantial differences between theory and experiment became apparent, for example for the metals Pb and Hg. The reason for this turned out to be that the assumption of constant $V$ is too simple: the electron phonon interaction is not instantaneous, but rather it is retarded in time, and the quasiparticle states have a finite lifetime. Eliashberg theory is an extension of BCS theory that explicitly includes the retardation effects \cite{eliashberg:1960}. It is beyond the scope of this Perspective to discuss the Eliashberg equations, which can be solved numerically. However, we introduce the key quantity of Eliashberg theory, the Eliashberg spectral function, which can be calculated (from first principles) or measured (by inverting tunnelling spectra). 

The Eliashberg spectral function, $\alpha^2F(\omega)$, is defined as:
\begin{equation}
 \alpha^2F(\omega) = \frac{1}{2\pi g(E_F)} \sum_{qj}\frac{\gamma_{qj}}{\omega_{qj}} \delta(\hbar\omega-\hbar\omega_{qj}),
\end{equation}
where the linewidth, $\gamma_{qj}$, of a phonon mode $j$ with a wave-vector $q$, $\omega_{qj}$, is given by:
\begin{equation}
\gamma_{qj} = 2\pi\omega_{qj}\sum_{knm}|g^j_{kn,k+qm}|^2\delta(\epsilon_{kn}-E_F)\delta(\epsilon_{k+qm}-E_F).
\label{eq:gamma}
\end{equation}
These equations describe the scattering of an electron on the Fermi surface with a resulting transfer of momentum to a phonon, where  $g^j_{kn,k+qm}$ is the electron phonon matrix element, or the probability, associated with this process. Given the Eliashberg function, the electron phonon coupling, $\lambda$, can be calculated via:
\begin{equation}
\lambda = 2 \int_0 ^ \infty \frac{\alpha^2F(\omega)}{\omega} d\omega .
\end{equation}

The Eliashberg formalism requires as input a parameter that describes the screened Coulomb repulsion between the electrons within a Cooper pair. This renormalized Coulomb repulsion parameter, $\mu^*$, is often called the Coulomb pseudopotential. Analytical expressions that can be used to approximate $\mu^*$ have been proposed, but typically it is used as a free parameter with reasonable values lying in the range of 0.1-0.2. We note that there is one theoretical approach, density functional theory for superconductors (SCDFT), which takes into account the retardation effects, and does not require any empirical parameters for the calculation of $T_c$ \cite{Luders:2005a,Marques:2005a}. However, only a few studies have used SCDFT for hydrides under high pressure, for example Refs.\ \cite{Flores:2016a,Flores:2016-P}.

One of the first simple expressions that could be used to estimate $T_c$ was developed by McMillan. Based on twenty-two numerical solutions of the Eliashberg equations for $0\le \mu ^* <0.25$,  $0<\lambda<1.5$, and a single shape for the $\alpha^2F(\omega)$ function modeled after the phonon DOS in Nb, McMillan showed that the following expression, where $\Theta_D$ is the Debye temperature, could be used to provide a reasonable estimate for $T_c$: \cite{McMillan:1968}
\begin{equation}
T_c = \frac{\Theta_D}{1.45}
\exp\left[-\frac{1.04(1+\lambda)}
  {\lambda-\mu^*(1+0.62\lambda)}\right] .
  \label{eq:mcmillan}
\end{equation}

Subsequently, Allen and Dynes carried out 200 numerical solutions for various shapes of the Eliashberg function, using a wide range of $\lambda$ values \cite{Allen:1975,Dynes:1972}. They proposed the following modification of the McMillan equation:
\begin{equation}
T_c = \frac{\omega_{\text{ln}}}{1.2}
\exp\left[-\frac{1.04(1+\lambda)}
  {\lambda-\mu^*(1+0.62\lambda)}\right],
  \label{eq:allen-dynes}
\end{equation}
where $\omega_{\text{ln}}$ is the logarithmic average frequency of the phonon modes obtained via
\begin{equation}
\omega_{\text{ln}}=\exp\left[\frac{2}{\lambda}\int_0^\infty\frac{d\omega}{\omega}\alpha^2 F(\omega)\ln\omega \right].
\end{equation}

Usually, once DFT calculations have been carried out to determine the Eliashberg spectral function, and from it $\lambda$, $T_c$ is estimated via Eq.\ \ref{eq:allen-dynes} for a range of $\mu^*$ values. However, whereas the Allen-Dynes modified McMillan equation implies that $T_c$ reaches a maximum limit when $\lambda \rightarrow \infty$, a maximum does not exist for the exact solution of the Eliashberg equations. In cases where the electron phonon coupling is strong, i.e.\ $\lambda \gtrsim$~1.5, Eq.\ \ref{eq:allen-dynes} often yields a lower limit to $T_c$. To remedy this, Allen and Dynes proposed multiplying Eq.\ \ref{eq:allen-dynes} by scaling factors that correct for the strong coupling, and shape dependence of $T_c$  \cite{Allen:1975}. However, for the studies discussed herein, $T_c$ was typically estimated solving the Eliashberg equations when $\lambda$ was large.

\section{Hydrogenic Motifs} \label{sec:motifs}

\subsection{Clathrate Based Hydrogenic Lattices} \label{sec:clathrate}
\begin{figure}[h!]
\begin{center}
\includegraphics[width=1\columnwidth]{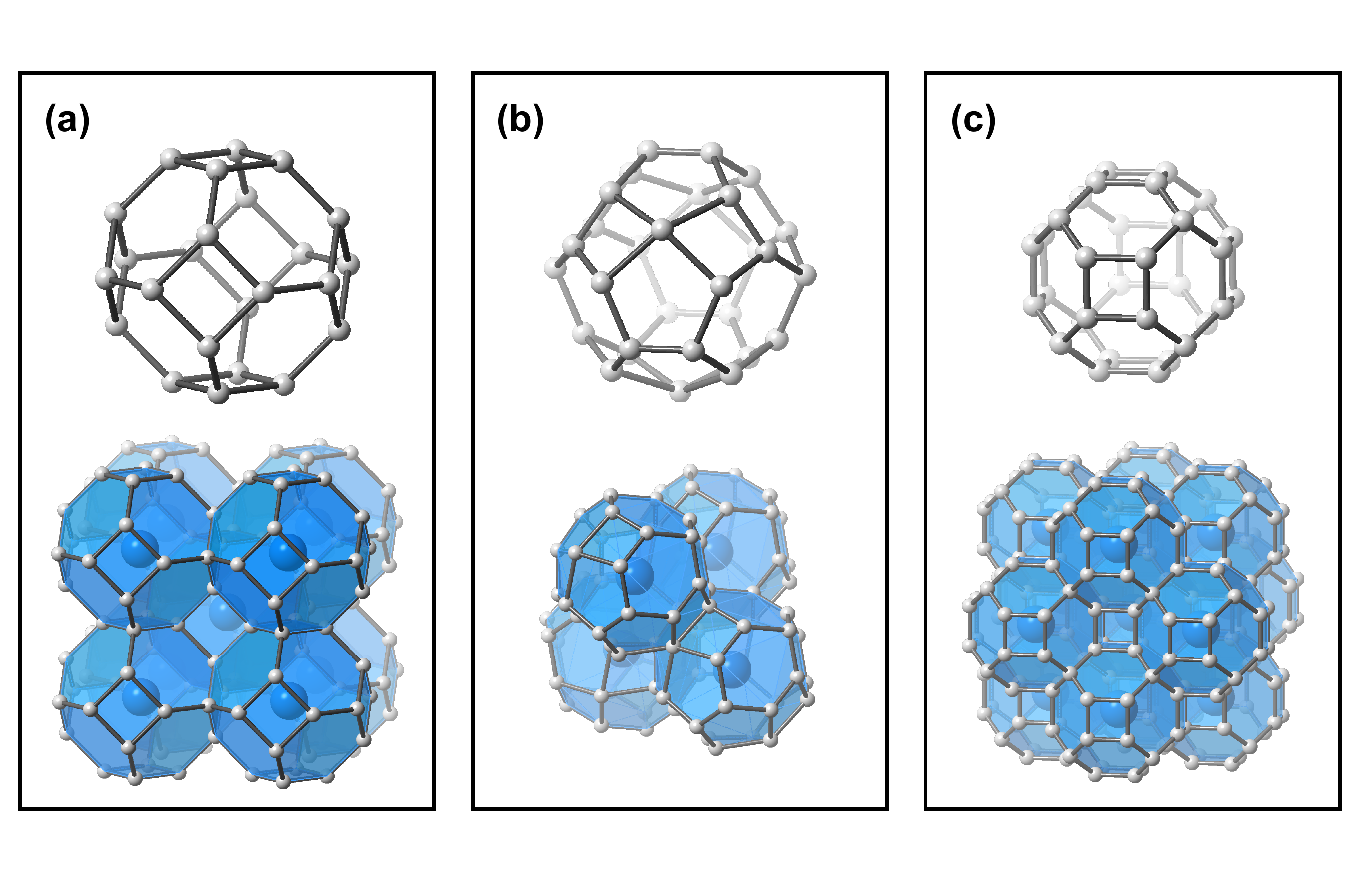}
\end{center}
\caption{(Top panel) The hydrogenic clathrate cage structures, and (bottom panel) extended: (a) $Im\bar{3}m$ MH$_6$, (b) $P6_3/mmc$ MH$_9$ and (c) $Fm\bar{3}m$ MH$_{10}$ systems they form with a subset of the alkaline earth and rare earth metals (see Table \ref{tab:clathrates}). White balls represent the hydrogen atoms, and the metal atoms are faint blue.}
\label{fig:Clathrate}
\end{figure}

A number of CSP-based studies have computed very high $T_c$s for binary alkaline earth or rare earth metal hydrides whose hydrogenic lattices are reminiscent of clathrates. The first system of this type to be predicted was a CaH$_6$ phase, which was thermodynamically and dynamically stable  above 150~GPa \cite{Wang:2012}. The building block of this phase is the  H$_{24}$ sodalite-like cage illustrated in the top panel of Fig.\ \ref{fig:Clathrate}(a), which consists of six square and eight hexagonal faces.  The metal atom lies in the center of this cage, and because the shortest H-H distance measures 1.24~\AA{}, these hydrogen atoms are only weakly bonded to each other. The crystalline lattice that is formed from joining the [4$^6$6$^8$] polyhedra at the faces, shown in the bottom panel of Fig.\ \ref{fig:Clathrate}(a), has $Im\bar{3}m$ symmetry. The band structure of $Im\bar{3}m$ CaH$_6$ contains degenerate bands at the $\Gamma$ point that are partially occupied. Therefore, Wang et al.\ suggested that a Jahn-Teller distortion, which removes this degeneracy, could yield a substantial electron phonon coupling parameter. Indeed, this phase was calculated to have a remarkably large $\lambda$ of 2.69, with the most significant contribution arising from modes associated with the vibrations of the atoms comprising the H$_4$ faces. Because of the large $\lambda$, $T_c$ was estimated by solving the Eliashberg equations. It  was calculated to be between 220-235~K for typical values of $\mu^*$,  as shown in Table \ref{tab:clathrates},  and it decreased at higher pressures.

\input{./table1.tex}

The prediction of high $T_c$ in CaH$_6$ inspired a computational study of an isotypic MgH$_6$ phase, which was found to become stable with respect to decomposition into MgH$_2$ and H$_2$ above 263~GPa when the ZPE was included \cite{Feng:2015a}. But, because a convex hull plotting the $\Delta H_F$ of $Im\bar{3}m$ MgH$_6$ along with the $\Delta H_F$ of other stoichiometries that were previously found to be stable at lower pressures  \cite{Lonie:2012} was not constructed, it is not clear if MgH$_6$ is thermodynamically stable or if it is only metastable. The estimated $T_c$ of MgH$_6$ at 300~K, see Table \ref{tab:clathrates}, was somewhat higher than the largest $T_c$ calculated for CaH$_6$. An isotypic strontium hexahydride was not a local minimum on the PES. However, $R\bar{3}m$ SrH$_6$, which can be derived from the $Im\bar{3}m$ structure by elongating four out of six of the closest metal-metal contacts and distorting the face that bisects them so it is no longer hexagonal, was stable at $\sim$250~GPa \cite{Hooper:2013,Wang:2015a}. This symmetry breaking distortion transformed the sodalite hydrogenic lattice into one-dimensional helical chains, and the resulting structure had a smaller $\lambda$, and concomitantly a lower $T_c$ \cite{Zurek:2018d} than its lighter cousins. A structurally analogous BaH$_6$ phase with a high $T_c$ could not be located via CSP  \cite{Hooper:2012b}. 

Moving to the $d$-block elements, computations have shown that ScH$_6$ \cite{Abe:Sc-2017,Qian:Sc-2017,Peng:Sc-2017,Zurek:2018b} and YH$_6$ \cite{Li:2015a,Liu:2017-La-Y,Peng:Sc-2017} phases with $Im\bar{3}m$ symmetry are also stable at pressures attainable in DACs. A related $I4/mmm$ symmetry ZrH$_6$ phase, which can be obtained via a distortion of the $Im\bar{3}m$ lattice, has also recently been found via CSP calculations \cite{Abe:2018-Zr}.  In one study \cite{Peng:Sc-2017} $Im\bar{3}m$  LaH$_6$  was computed to be thermodynamically stable at 100~GPa and 150~GPa, whereas in another investigation \cite{Liu:2017-La-Y} the LaH$_6$ stoichiometry was predicted to lie slightly above the convex hull, and assume the same $R\bar{3}m$ symmetry structure previously found for SrH$_6$. As shown in Table \ref{tab:clathrates}, the electron phonon coupling of YH$_6$ was computed to be particularly high, $\lambda=2.93$, and therefore its $T_c$ was estimated to be at least 100~K higher than any of the other stable superconducting $d$-block hexahydrides. 

The [4$^6$5$^6$6$^6$] H$_{29}$ polyhedron shown in the top panel of Fig.\ \ref{fig:Clathrate}(b) is comprised of six irregular squares, six pentagons and six hexagons. It is the building block of a hydrogenic clathrate-like lattice with $P6_3/mmc$ symmetry that has been predicted for a number of MH$_9$ stoichiometry rare-earth hydrides \cite{Peng:Sc-2017}. Phases with the MH$_{10}$ stoichiometry, on the other hand, were often found to assume an $Fm\bar{3}m$ symmetry structure comprised of the H$_{32}$ [4$^6$6$^{12}$] polyhedron, which is shown in Fig.\ \ref{fig:Clathrate}(c) \cite{Peng:Sc-2017,Liu:2017-La-Y}. In the clathrate and zeolite community the latter is known as `AST' \cite{Amri:2010a}.  The stability of the clathrate cages under pressure was attributed to their dense packing, which yields a substantially lower $PV$ term to the enthalpy. For example, at 300~GPa a $P6_3/mmc$ symmetry YH$_9$ clathratic phase had the smallest volume out of any of the 500 lowest enthalpy structures found via CSP searches \cite{Peng:Sc-2017}. In some cases the enthalpy of a phase that did not possess a clathrate-like structure was slightly lower, but the ZPE-corrected enthalpies of the two dynamically stable phases were within a few meV/atom of each other \cite{Peng:Sc-2017}. However, because the estimated $T_c$s of the clathrate-like structures were significantly higher, they are provided in Table \ref{tab:clathrates}. Below 210~GPa, $Fm\bar{3}m$ LaH$_{10}$ was not a minimum on the PES, but a $C2/m$ symmetry lattice that could be derived from the ideal structure via a slight distortion was dynamically stable \cite{Liu-2018-La}. 

As shown in Table \ref{tab:clathrates}, enticingly high $T_c$ values, some of which even surpass room temperature, have been computed for the YH$_9$, YH$_{10}$ and LaH$_{10}$ stoichiometries. Although stable clathritic LaH$_9$, CeH$_9$, CeH$_{10}$, and PrH$_9$ phases were also found via CSP, their computed $T_c$ values were much lower, $<$~56~K \cite{Peng:Sc-2017}, because the lower frequency vibrations, which are a result of the heavier element, reduce $T_c$ in a BCS superconductor. On the other hand, a metastable $R\bar{3}m$ symmetry AcH$_{10}$ phase was computed to have a quite high $T_c$ \cite{Semenok-2018}. It's hydrogenic lattice resembled the clathrate structure illustrated in Fig.\ \ref{fig:Clathrate}(c), except the square faces were replaced by trapezoids. Out of any of the binary hydrides studied computationally YH$_9$, YH$_{10}$ and LaH$_{10}$ hold the record $T_c$ values to date, as seen in Fig.\  \ref{fig:periodic-table}.

The aforementioned theoretical predictions inspired experimental studies. Last year a superhydride of lanthanum consistent with the theoretically predicted structure for LaH$_{10}$ was synthesized at 170~GPa, and decompression led to a $Fm\bar{3}m \rightarrow R\bar{3}m \rightarrow C2/m$ phase transformation  \cite{Geballe:2018a}. And, two very recent studies observed dramatic drops in resistivity in the lanthanum/hydrogen system under pressure, tantalizing with the allure of high temperature superconductivity \cite{Zulu:2018-La,Drozdov:2018-La}. If confirmed, both of these findings would break the current record for the highest measured $T_c$ values previously observed in the hydrogen/sulfur system \cite{Drozdov:2015a}. Somayzaulu et al.\ measured a $T_c$ of between 245-280~K at 190-200~GPa in a sample whose diffraction peaks were consistent with the previously synthesized LaH$_{10\pm x}$ phase \cite{Zulu:2018-La}. As Table \ref{tab:clathrates} reveals, this value is in good agreement with theoretical estimates for both the $Fm\bar{3}m$ and the $C2/m$ symmetry LaH$_{10}$ phases. Drozdov and co-workers, who employed a different experimental technique to synthesize a so-far uncharacterized superhydride of lanthanum, reported a maximum $T_c$ of 215~K at 150~GPa \cite{Drozdov:2018-La}. Because the samples were synthesized using different methods, and the measurements were obtained at different pressures it is likely that the phase, or mixture of phases, made in the two studies are not the same.

\subsection{Mixed Molecular and Atomic Hydrogen} \label{sec:mixed}

Many of the alkali metal, alkaline earth and rare earth polyhydrides that are stable under pressure possess both atomic and molecular hydrogen. To better understand the electronic structure of phases containing these structural motifs, let us assume a full transfer of the metals' valence electrons to hydrogen, yielding  H$^-$ atoms, and H$_2^{\delta-}$ molecules. Phases with $\delta=0$ are expected to metallize as a result of pressure induced broadening of the H$^-$ and H$_2$ $\sigma^*$ anti-bonding based bands \cite{Zurek:2009c}. As a result, they generally do not have a high DOS at $E_F$, and they are therefore not expected to be good superconductors \cite{Zurek:2016d,Shamp:2016}. Phases with $\delta > 0$, on the other hand, are metallic because of partial filling of the H$_2$ $\sigma^*$-bands. As a result, they are likely to be good metals with the potential for high temperature superconductivity.

\begin{figure}[h!]
\begin{center}
\includegraphics[width=1\columnwidth]{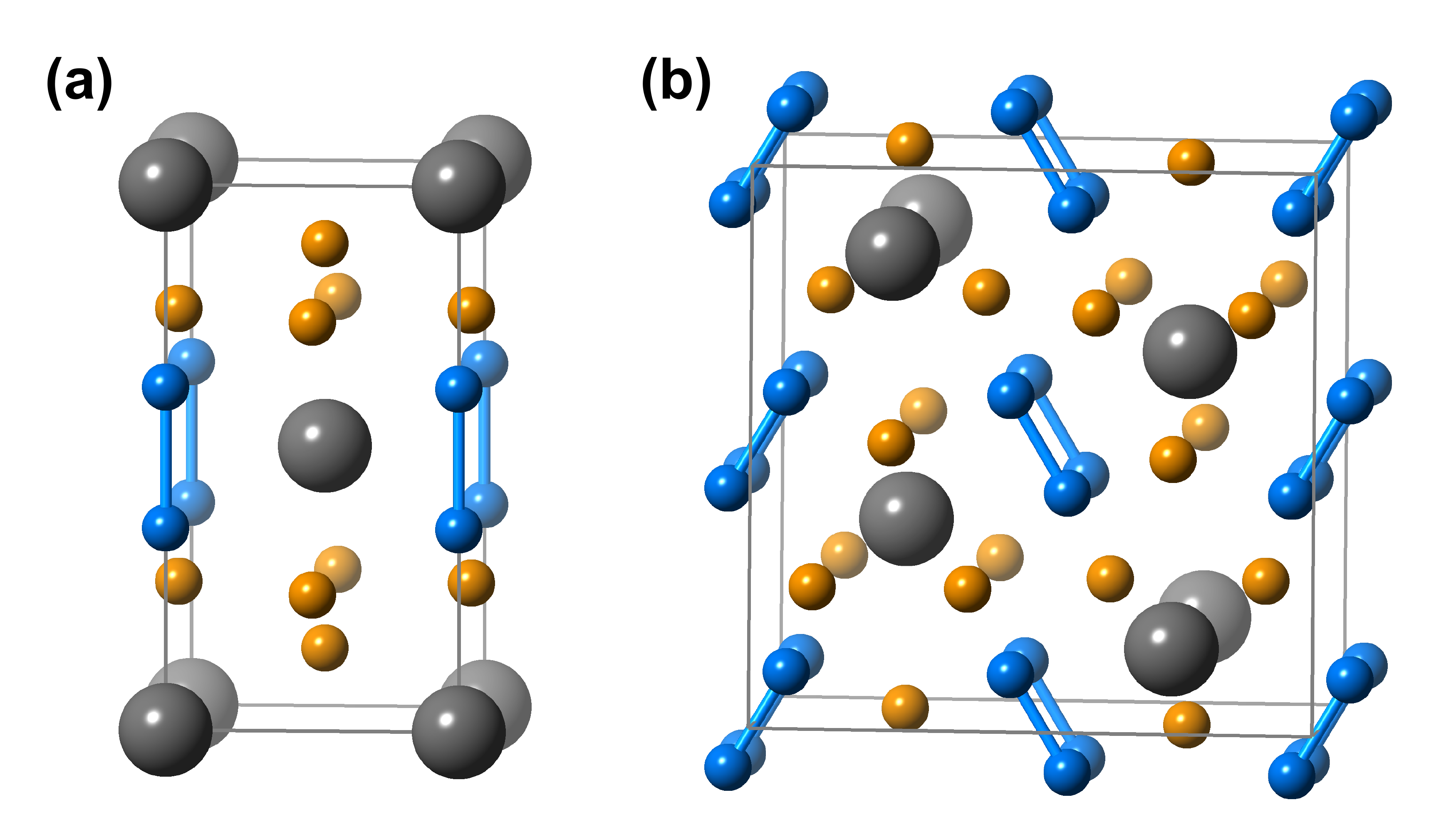}
\end{center}
\caption{(a) The $I4/mmm$ symmetry MH$_4$ structure that has been predicted to be stable for many of the alkaline earth and rare earth hydrides under pressure. (b) The crystal structure of $Cmcm$ ScH$_7$ at 300~GPa \cite{Zurek:2018b}. Grey balls represent the metal atoms; white balls represent H$^-$ atoms; H$_2^{\delta -}$ units are colored in blue.}
\label{fig:mixed}
\end{figure}

One example of the former is the $I4/mmm$ symmetry CaH$_4$ (or Ca$^{2+}$(H$^-$)$_2$H$_2$) structure illustrated in Fig.\ \ref{fig:mixed}(a), which was first predicted via CSP \cite{Wang:2012}, and recently synthesized in a laser heated DAC \cite{Zurek:2018c}. 
At 120~GPa a Bader analysis, which typically underestimates the charge transfer, yielded values of H$^{-0.5}$ and H$_2^{-0.05}$.
The H-H bond length at 120~GPa, 0.81~\AA{}, was found to be 33\% larger than in H$_2$ at the same pressure. Theoretical calculations revealed that the weakening of the H-H bond in CaH$_4$ resembles the mechanism leading to anomalously long H-H bonds in Kubas-like molecular complexes wherein an H$_2$ molecule binds side-on to the metal center \cite{Zurek:2018c}. In both the molecular and the high pressure systems donation of $\sigma$ electrons from H$_2$ to a vacant metal d-orbital, and d~$\rightarrow \sigma^*$ back-donation weaken the H-H bond. Because both the PBE-GGA and hybrid HSE06 functionals predicted that $I4/mmm$ CaH$_4$ would be a weak metal at this pressure, its $T_c$ was not computed. 

CSP calculations have also shown that isotypic SrH$_4$ \cite{Hooper:2013,Wang:2015a} and MgH$_4$ \cite{Abe:2018-Zr} phases are stable under pressure. Whereas $I4/mmm$ SrH$_4$ did not metallize within its range of stability, MgH$_4$ did, at least within PBE. Table \ref{tab:mixed} shows that the $T_c$ of MgH$_4$ at 255~GPa was estimated to be close to the boiling point of liquid nitrogen . However, because this system metallizes as a result of pressure induced band broadening, PBE likely overestimates its DOS at $E_F$, and therefore its $T_c$. A BaH$_4$ stoichiometry did not lie on the convex hull \cite{Hooper:2012b}.

DFT-based CSP searches have predicted that this same $I4/mmm$ symmetry structure will be stable for many of the rare earth metal tetrahydrides under pressure: ScH$_4$ \cite{Abe:Sc-2017,Qian:Sc-2017,Zurek:2018b,Peng:Sc-2017},  YH$_4$ \cite{Peng:Sc-2017}, LaH$_4$ \cite{Liu:2017-La-Y}, CeH$_4$ \cite{Peng:Sc-2017} and PrH$_4$ \cite{Peng:Sc-2017}. Moreover, theoretical studies that compared the enthalpies of a set of candidate structures, which were chosen based upon structural analogy, suggest that isotypic NdH$_4$, PmH$_4$, SmH$_4$, EuH$_4$, GdH$_4$, TbH$_4$, DyH$_4$, HoH$_4$, ErH$_4$, TmH$_4$, and LuH$_4$ phases lie on the convex hull at experimentally attainable pressures \cite{Peng:Sc-2017}. CSP searches also predicted that ZrH$_4$ would assume this structure \cite{Abe:2018-Zr}.

Out of all of the stable $d$-block $I4/mmm$ symmetry tetrahydrides, only ScH$_4$ and YH$_4$ were estimated to have $T_c$s above the boiling point of liquid nitrogen, see Table \ref{tab:mixed}.  Assuming a +3 oxidation state for the metal atom yields a formula of M$^{3+}$(H$^-$)$_2$H$_2^-$ for these systems. Electron transfer into the H$_2$ $\sigma^*$-bands yields a high DOS at $E_F$, as expected. The nearest neighbor H-H distances measured 1.21~\AA{} at 250~GPa in ScH$_4$ \cite{Zurek:2018b}, and 1.39~\AA{} at 120~GPa in YH$_4$ \cite{Li:2015a}. The long intramolecular H-H bond lengths, which are conducive to a large electron phonon coupling, have two origins. First, because these are $d$-elements the Kubas-like mechanism responsible for the elongation of the H$_2$ bond within $I4/mmm$ CaH$_4$ is likely to be important. Secondly, charge is donated from the electropositive element into the H$_2$ $\sigma^*$-bands, thereby weakening and lengthening the H-H bond.

An isotypic ZrH$_4$ phase was estimated to have a $T_c$ of 47~K at 230~GPa \cite{Abe:2018-Zr}. At this pressure the nearest neighbor H-H distance, 1.21~\AA{}, and electron phonon coupling constant, $\lambda=0.89$, was similar to the values computed for ScH$_4$ and YH$_4$. The $T_c$ of an $I4/mmm$ symmetry LaH$_4$ phase was estimated to be 5-10~K at 300~GPa \cite{Liu:2017-La-Y}. Its decreased $T_c$ as compared to ScH$_4$ and YH$_4$ is likely a result of the lower frequency vibrations, which stem from the presence of the heavier metal atom. For similar reasons, the heavier $I4/mmm$ symmetry tetrahydrides are not expected to be superconducting at high temperatures, and their $T_c$s were therefore not computed \cite{Peng:Sc-2017}. 
%
\input{table2.tex}

Another phase with mixed molecular and atomic hydrogen that was predicted to be superconducting at high temperatures is the $Cmcm$ ScH$_7$ structure shown in Fig.\ \ref{fig:mixed}(b) \cite{Zurek:2018b}. Because this system contains three atomic and two molecular hydrogen atoms per scandium atom, at first glance one would guess that its chemical formula can be written as Sc$^{3+}$(H$^-$)$_3$(H$_2$)$_2$. At 300~GPa the H-H bond, which measures 0.956~\AA{}, is significantly elongated relative to that of molecular H$_2$ at this pressure. In contrast to the results for CaH$_4$, the computed average Bader charges on the ``atomic'' and ``molecular'' hydrogen atoms were nearly identical, -0.186$e$ and -0.138$e$, respectively, suggesting that the formula given above does not adequately describe the electronic structure of $Cmcm$ ScH$_7$. Moreover, the substantial charge transfer to molecular hydrogen is consistent with its long H-H bond, the high DOS at $E_F$, and the large electron phonon coupling, which ultimately both yield a sizeable $T_c$.

\subsection{Other Unique Hydrogenic Motifs} \label{sec:other}

In addition to the aforementioned systems whose hydrogenic lattices were either clathrate-like or possessed molecular and atomic hydrogen, four more phases, with quite unusual hydrogenic motifs, were computed to be superconducting up to high temperatures. Comparison of Tables \ref{tab:clathrates} and \ref{tab:mixed} with Table \ref{tab:other} reveals that their $T_c$ values were generally intermediate to the other two classes of systems described above.  

\begin{figure}[h!]
\begin{center}
\includegraphics[width=0.9\columnwidth]{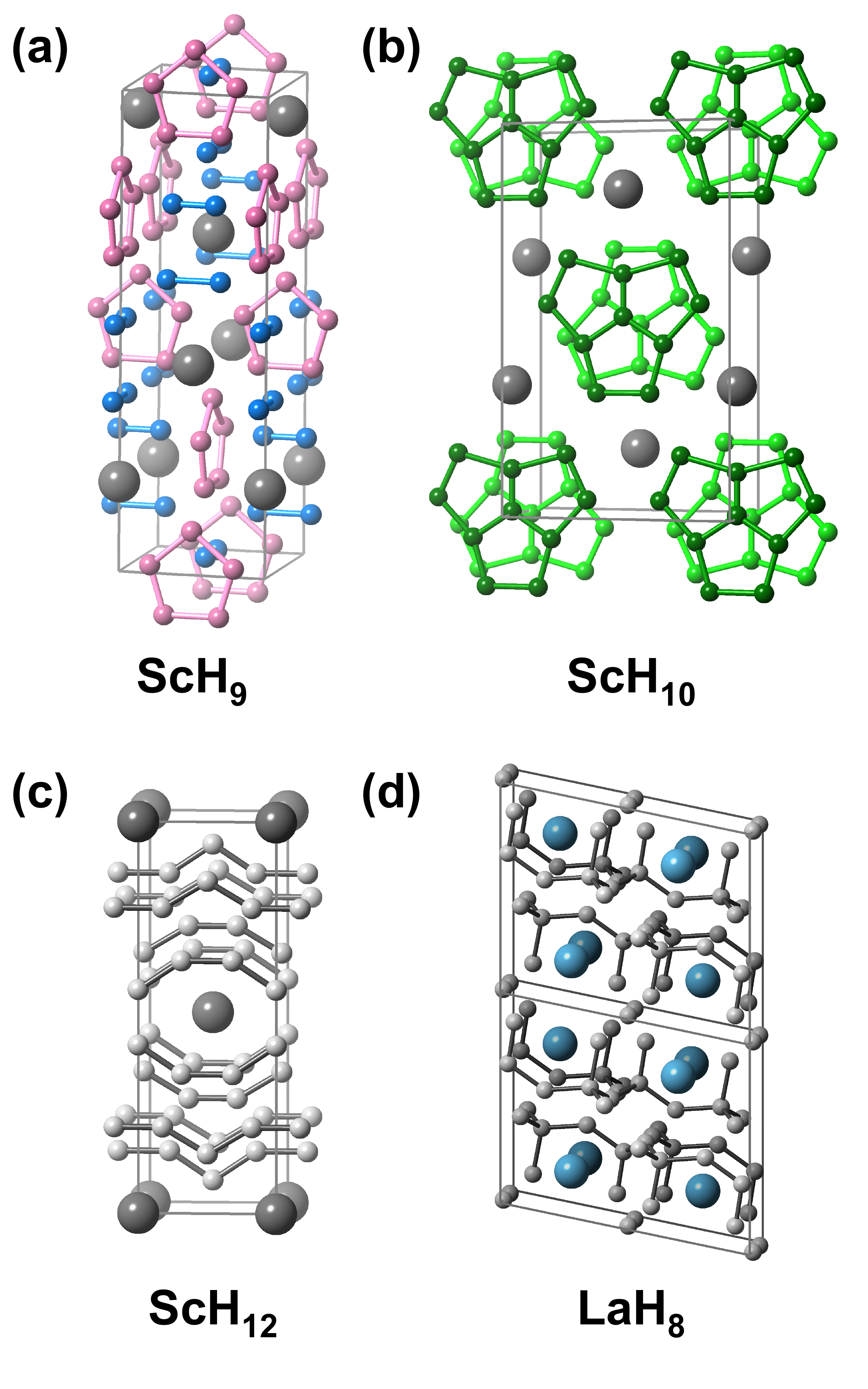}
\end{center}
\caption{The crystal structures of: (a) $I4_1md$ ScH$_9$ at 300~GPa, (b) $Cmcm$ ScH$_{10}$ at 250~GPa and (c) $Immm$ ScH$_{12}$ at 350~GPa \cite{Zurek:2018b}. Grey balls represent the scandium atoms; white balls represent the hydrogen atoms; H$_5^{\delta -}$ units are colored in pink; H$_2^{\delta -}$ units are colored in blue; H$_{10}^{\delta -}$ units are colored in green.}
\label{fig:ScHn}
\end{figure}

$I4_1md$ ScH$_9$, shown in Fig.\ \ref{fig:ScHn}(a), was predicted to be stable in a narrow pressure range of 285-325~GPa when the ZPE was included \cite{Zurek:2018b}. This phase contained 1-dimensional chains of edge-sharing H$_5^{\delta-}$ pentagons, as well as molecular H$_2^{\delta-}$ units, with the shortest H-H distance being 0.90~\AA{} at 300~GPa. The hydrogenic lattice of a $Cmcm$ ScH$_{10}$ phase \cite{Peng:Sc-2017,Zurek:2018b} consisted solely of these same ``H$_5$'' pentagonal motifs, their edges shared so as to form discrete H$_{10}^{\delta-}$ molecular units, as illustrated in Fig.\ \ref{fig:ScHn}(b). One way to describe the hydrogenic lattice of the $Immm$ ScH$_{12}$ structure shown in Fig.\ \ref{fig:ScHn}(c) would be of H$_8^{\delta-}$ ``octagons'' and H$_5^{\delta-}$ ``pentagons'', where six edges of each octagon are connected by pentagons and the other edges opposite each other are connected by another octagon, thereby forming a 1-dimensional framework. The ZPE-corrected enthalpies at 0~K predicted that ScH$_{10}$ and ScH$_{12}$ would be thermodynamically stable between 220-285~GPa and above 325~GPa, respectively \cite{Zurek:2018b}.

Finally, the LaH$_8$ stoichiometry was found to lie on the 150~GPa and 300~GPa convex hulls \cite{Liu:2017-La-Y}. At 300~GPa the 2-dimensional hydrogenic lattice of this $C2/c$ symmetry phase, shown in Fig.\ \ref{fig:ScHn}(d), consisted of edge-sharing puckered dodecagons wherein every second hydrogen atom was capped so that it became four-fold coordinate with H-H-H angles that deviated somewhat from those found in a perfect tetrahedron, 104.7$^\circ$-116.3$^\circ$. In addition, this phase contained hydrogen atoms, which did not appear to be bonded to any other hydrogen atoms, that fell between the puckered layers.  

\input{table3.tex}

Even though a $P4/nmm$ BeH$_2$ phase was predicted to have a $T_c$ of 97.0~K at 365~GPa, it is not discussed here since the hydrogen atoms within it were not bonded to each other \cite{Yu:2014}. CSP searches showed that BeH$_n$ with $n>2$ are not thermodynamically stable with respect to decomposition into the classic BeH$_2$ hydride and H$_2$ up to 200~GPa \cite{Hooper:2012b}. We also mention that an $Fddd$ symmetry ZrH$_4$ phase that was also only compromised of atomistic hydrogen (nearest neighbor H-H distance of 1.38~\AA{}) was predicted to have a $T_c$ of 78.0~K at 140~GPa \cite{Abe:2018-Zr}. And that quite high $T_c$ values, ranging from $\sim$100-225~K, have also been predicted for some actinide polyhydrides with interesting structural motifs, for example $C2/m$ AcH$_8$ and $P\bar{6}m2$ AcH$_{16}$, which become stable above $\sim$100~GPa, or a metastable $I4/mmm$ AcH$_{12}$ phase \cite{Semenok-2018}.

\section{Outlook} \label{sec:outlook}
Crystal structure prediction techniques coupled with first-principles calculations have found a number of superconducting hydride phases that are either stable or metastable at pressures that can be achieved within diamond anvil cells. Binary hydrides with remarkably high estimated superconducting critical temperatures, $T_c$s, were predicted within two regions of the periodic table: the $p$-block elements and the alkaline earth and rare earths. Superconductivity up to 203~K was experimentally confirmed in the first region within a sample of hydrogen sulfide that was compressed to 150~GPa \cite{Drozdov:2015a}. This is the system with the record highest $T_c$ confirmed to date. Recent experiments measuring $T_c$s of 215~K \cite{Drozdov:2018-La} or 260-280~K \cite{Zulu:2018-La} in the lanthanum/hydrogen system under pressure suggest that the $T_c$ of hydride phases in the second region may even surpass the record attributed to H$_3$S, and tempt with the lure of room temperature superconductivity. 

Experimental techniques to synthesize high hydride phases under pressures have made remarkable advances in the last years \cite{Zurek:2018d}. We therefore expect the experimental search for superconductivity in these systems will accelerate, with the predictions from theory used as a guide. Given what has come before \cite{Zurek:2014i}, it is likely that a feedback loop between theory and experiment will be required to characterize the phases that have been made, and to gain an understanding of their electronic structure and properties. 

The reason why hydrides with clathrate-based hydrogenic lattices are estimated to have such high $T_c$ values can be traced back to their large density of states at the Fermi level,  which is derived primarily from hydrogen-like states, and the pronounced impact on the electronic structure that results from the motions of the atoms comprising their hydrogenic lattices. It has been suggested that perturbing hydrogen atoms within quasi-molecular units such as H$_2^{\delta-}$ would not have such a large impact on the electronic structure, thereby resulting in a smaller  electron-phonon coupling, with concomitantly lower $T_c$ values \cite{Shamp:2016,Zhang:2017}. In accordance with this reasoning, the predicted $T_c$s for the MH$_4$ phases containing mixed molecular and atomic hydrogen are generally lower than those of the clathrate-like systems as shown in this Perspective. 

However, this cannot be the full story because quite high $T_c$ values have been calculated for a few of the scandium polyhydrides discussed herein whose hydrogenic lattices contain quasi-molecular units. Moreover, the factors driving the stability of these phases are not well understood, and it remains a grand challenge to find ways to quench these systems to pressures that can be easily achieved industrially. There is still much to be learned in this exciting field of research.

\section*{Acknowledgments}
We acknowledge the NSF (DMR-1505817) for financial, and the Center for Computational Research (CCR) at SUNY Buffalo for computational support. T.\ B.\ thanks the US Department of Energy, National Nuclear Security Administration, through the Capital-DOE Alliance Center under Cooperative Agreement DE-NA0003858 for financial support. We thank Prof.\ Yansun Yao from the University of Saskatchewan and Dr.\ Peng Feng from Luoyang Normal University for useful comments.

%\bibliography{Review}


\end{document}

%% file: table1.tex
%\begin{longtable}{c c c c c c c c c}
\begin{table}[h!]
\begin{center}
\caption{Electron phonon coupling parameter, $\lambda$, and superconducting critical temperature, $T_c$, of phases assuming the $Im\bar{3}m$ (MH$_6$), $P6_3/mmc$ (MH$_9$) and $Fm\bar{3}m$ (MH$_{10}$) sodalite structures illustrated in Fig.\ \ref{fig:Clathrate} computed at the pressures, $P$, and using the values of the Coulomb pseudopotential, $\mu^*$, listed. Data is also provided for three structurally related phases that can be derived via a symmetry breaking distortion of the clathrate cages. } 
\begin{tabular}{ c | c | c | c | c }
\hline 
\hline
System       & $P$ (GPa) & $\lambda$ &  $\mu ^*$   &   $T_c$ (K) \\
\hline
 MgH$_6$      & 300 & 3.29 & 0.12    & 263 $^{a,}$ \cite{Feng:2015a} \\
 CaH$_6$      & 150 & 2.69 &0.13-0.10    & 220-235 $^{b,}$ \cite{Wang:2012} \\
 SrH$_6^*$      & 250 & 1.10 & 0.10   & 156 $^{b,}$ \cite{Zurek:2018d} \\
 ScH$_6$      & 350 & 1.25 & 0.10 & 135 $^{a,}$ \cite{Zurek:2018b}, 169 $^{b,}$ \cite{Zurek:2018b} \\
 ScH$_6$	 & 285 & 1.33 & 0.13-0.10 & 130-147 $^{a,}$ \cite{Abe:Sc-2017} \\
 ScH$_6^{\dagger}$	 & 300 & 1.20 & 0.13-0.10 & 90-100 $^{b,c,}$ \cite{Peng:Sc-2017} \\
 YH$_6$      & 120 & 2.93 & 0.13-0.10 & 251-264 $^{b,}$ \cite{Li:2015a,Peng:Sc-2017} \\
ZrH$_6^{\P}$      & 295 & 1.20 & 0.13 & 114 $^{b,}$ \cite{Abe:2018-Zr} \\
 LaH$_6$      & 100 & 2.00 & 0.13-0.10 & 150-160 $^{b,c,}$ \cite{Peng:Sc-2017} \\
 \hline
 ScH$_9$      & 400 & 1.50 & 0.13-0.10 & 150-190 $^{b,c,}$ \cite{Peng:Sc-2017} \\
 YH$_9^{\dagger}$    & 150 & 4.42 & 0.13-0.10 & 276-253 $^{b,}$ \cite{Peng:Sc-2017} \\
\hline
 YH$_{10}$      & 400 & 2.41 & 0.13-0.10 & 287-303 $^{b,}$ \cite{Peng:Sc-2017} \\
 YH$_{10}$               & 250 & 2.58 & 0.13-0.10 & 244-265 $^{a,} \cite{Liu:2017-La-Y}$, 305-326 $^{b,}$ \cite{Liu:2017-La-Y} \\
 LaH$_{10}^{\S}$      & 200 & 2.28 & 0.10 & 288 $^{a,}$ \cite{Peng:Sc-2017} \\
 LaH$_{10}$      & 210 & 3.41 & 0.13-0.10 & 219-238 $^{a,} \cite{Liu:2017-La-Y}$, 274-286 $^{b,}$ \cite{Liu:2017-La-Y} \\
LaH$_{10}^{\ddagger}$      & 200 & 3.57 & 0.13-0.10 & 218-200 $^{a,}$ \cite{Liu-2018-La}, 245-229 $^{b,}$ \cite{Liu-2018-La} \\
\hline
AcH$_{10}^{\textdollar}$ & 200 & 3.46 & 0.15-0.10 & 177-204.1 $^{a,}$ \cite{Semenok-2018}, 226-251 $^{b,}$ \cite{Semenok-2018} \\
% CeH$_{10}$      & 200 & 1.00 & 0.13-0.10 & 50-60 $^{a,c,}$ \cite{Peng:Sc-2017} \\
\hline
\hline
\end{tabular}
%\end{longtable}
\label{tab:clathrates}
\end{center}
\noindent 
$^a$ $T_c$ was calculated using the Allen-Dynes modified McMillan equation,  Eq.\ \ref{eq:allen-dynes}. \\
$^b$ $T_c$ was calculated by solving the Eliashberg equations numerically. \\
%$^c$ $T_c$ was calculated using the simplified Allen-Dynes formula. \\ 
$^c$ Values were estimated from plots in the original papers. \\
$^*$ This $R\bar{3}m$ symmetry phase can be derived by distorting the $Im\bar{3}m$ structure. \\
$^{\dagger}$ At this pressure the enthalpy of a different phase was slightly lower than for the clathrate structure whose $T_c$ is provided. \\
$^{\P}$ This $I4/mmm$ symmetry phase can be derived by distorting the $Im\bar{3}m$ structure. \\
$^{\S}$ This stoichiometry did not lie on the convex hull without ZPE at 0~K. \\
$^{\ddagger}$ This $C2/m$ symmetry phase can be derived by distorting the $Fm\bar{3}m$ structure. \\
$^{\textdollar}$ This $R\bar{3}m$ symmetry phase can be derived by distorting the $Fm\bar{3}m$ structure.
\end{table}

%% file: table2.tex
%\begin{longtable}{c c c c c c c c c}
\begin{table}[h!]
\begin{center}
\caption{Electron phonon coupling parameter, $\lambda$, and superconducting critical temperature, $T_c$, of phases containing mixed molecular and atomic hydrogen computed at the pressures, $P$, and using the values of the Coulomb pseudopotential, $\mu^*$, listed. Data is provided only for systems whose computed $T_c$s are higher than the boiling point of liquid nitrogen. MgH$_4$, ScH$_4$ and YH$_4$ adopt the structure illustrated in  Fig.\ \ref{fig:mixed}(a). ScH$_7$ is shown in Fig.\ \ref{fig:mixed}(b).} 
\begin{tabular}{ c | c | c | c | c }
\hline 
\hline
System       & $P$ (GPa) & $\lambda$ &  $\mu ^*$   &   $T_c$ (K) \\
\hline
  MgH$_4$      & 255 & 0.88 & 0.13 & 81 $^{b,}$ \cite{Abe:2018-Zr} \\
 ScH$_4$	     & 120   & 1.68 & 0.10 & 92 $^{a,}$\cite{Zurek:2018b}, 163 $^{b,}$ \cite{Zurek:2018b} \\
 			 & 195  &  0.89 & 0.13-0.10 & 67-81 $^{a,}$ \cite{Abe:Sc-2017} \\
 			 & 200 & 0.99 & 0.10 & 98 $^{a,}$ \cite{Qian:Sc-2017} \\ 
 YH$_4$      & 120   & 1.01 & 0.13-0.10 & 84-95 $^{b,}$ \cite{Li:2015a} \\
\hline
 ScH$_7$     & 300 & 1.84 & 0.10 & 169 $^{a,}$\cite{Zurek:2018b}, 213 $^{b,}$ \cite{Zurek:2018b} \\
\hline
\end{tabular}
%\end{longtable}
\label{tab:mixed}
\end{center}
\noindent 
$^a$ $T_c$ was calculated using the Allen-Dynes modified McMillan equation,  Eq.\ \ref{eq:allen-dynes}. \\ 
$^b$ $T_c$ was calculated by solving the Eliashberg equations numerically. \\
%$^c$ $T_c$ was calculated using the simplified Allen-Dynes formula. \\
%$^c$ Values were estimated from plots found in the original papers. \\
\end{table}

%% file: table3.tex
%\begin{longtable}{c c c c c c c c c}
\begin{table}[h!]
\begin{center}
\caption{Electron phonon coupling parameter, $\lambda$, and superconducting critical temperature, $T_c$, of the $I4_{1}md$ ScH$_9$, $Cmcm$ ScH$_{10}$, $Immm$ ScH$_{12}$, and $C2/m$ LaH$_8$ phases illustrated in Fig.\ \ref{fig:ScHn}, computed at the pressures, $P$, and using the values of the Coulomb pseudopotential, $\mu^*$, provided.} 
\begin{tabular}{ c | c | c | c | c }
\hline 
\hline
System       & $P$ (GPa) & $\lambda$ &  $\mu ^*$   &   $T_c$ (K) \\
\hline
$I4_{1}md$ ScH$_9$ & 300 & 1.94 & 0.10 & 163 $^{a,}$ \cite{Zurek:2018b}, 233 $^{b,}$ \cite{Zurek:2018b} \\
$Cmcm$ ScH$_{10}$ & 250 & 1.17 & 0.10 & 120 $^{a,}$ \cite{Zurek:2018b}, 143 $^{b,}$ \cite{Zurek:2018b} \\
$Immm$ ScH$_{12}$ & 350 & 1.23 & 0.10 & 141 $^{a,}$ \cite{Zurek:2018b}, 194 $^{b,}$ \cite{Zurek:2018b} \\
$C2/m$ LaH$_8$ & 300 & 1.12 & 0.13-0.10 & 114-131 $^{a,}$ \cite{Liu:2017-La-Y}, 138-150 $^{b,}$ \cite{Liu:2017-La-Y} \\
%$C2/m$ SnH$_{12}$ & 250 & 1.250 & 0.13-0.10 & 83-93 $^{b,}$ \edit{Esfahani:2016} \\
%Unique linear H4- units. Although SnH14 has high Tc, it only have H3- units.
\hline
\hline
\end{tabular}
%\end{longtable}
\label{tab:other}
\end{center}
\noindent 
$^a$ $T_c$ was calculated using the Allen-Dynes modified McMillan equation, Eq.\ \ref{eq:allen-dynes}. \\ 
$^b$ $T_c$ was calculated by solving the Eliashberg equations numerically. \\
%$^c$ $T_c$ was calculated using the simplified Allen-Dynes formula. \\
%$^c$ Values were estimated from plots found in the original papers. \\
\end{table}

%LiH6: only have H2 unit
%TaH6: only have H unit
%B2H6: only have H (3-center-2-electron)
%AlH3(H2): only have H2 and H
%GaH3: only have H unit
%Si2H6: only have H unit
%GeH3: only have H unit
%PbH4(H2)2: only have H2 unit
%AsH8: only have quasi-H2 unit
%SbH4: only have quasi-H2 unit
%BiH5: only have H3- unit
%H4Te: only have quasi-H2 unit

%% file: zurek-spotlight.bbl
\begin{thebibliography}{189}
\bibitem [{not({\natexlab{b}})}]{note2}%
  \BibitemOpen
  \href@noop {} {} \ \bibinfo {note} {The high $T_c$ values
  predicted for FeH$_5$ \cite{Majumdar:2017a,Kvashnin:2018a} are not included,
  since subsequent studies \cite{Heil:2018a,Zurek:2018h} have questioned these
  results}\BibitemShut {NoStop}%
\end{thebibliography}
